\documentclass[aps, secnumarabic, superscriptaddress]{revtex4-2}
\usepackage{amsmath,amssymb}
\usepackage{graphicx}
\usepackage{newtxtext,newtxmath}

\begin{document}

\title{Global Phase Locking in Superradiant Chorus Emissions}

\author{Xin Tao}
\email[]{xtao@ustc.edu.cn}
\affiliation{Department of Geophysics and Planetary Sciences, University of Science and Technology of China, Hefei, China}

\author{Fulvio Zonca}
\affiliation{Center for Nonlinear Plasma Science and C.R. ENEA Frascati, C.P. 65, 00044 Frascati, Italy}
\affiliation{Institute for Advanced Study in Physics, Zhejiang University, China}

\author{Liu Chen}
\affiliation{Center for Nonlinear Plasma Science and C.R. ENEA Frascati, C.P. 65, 00044 Frascati, Italy}
\affiliation{Institute of Fusion Theory and Simulation, Zhejiang University, China}

\date{\today}

\begin{abstract}
  Coherent wave amplification in inhomogeneous systems is usually described in
  terms of resonance and nonlinear phase locking, but these local phase
  conditions do not determine how the wave amplitude evolves when the resonance
  condition changes during wave propagation. Here we identify a global
  phase-locking principle, in which the stable resonant phase remains nearly
  unchanged as the resonance detuning evolves. This condition requires
  the wave amplitude to track the evolving detuning and leads to linear
  amplitude growth with distance, producing the quadratic intensity scaling
  characteristic of superradiant emission. We demonstrate this mechanism using
  self-consistent kinetic simulations of whistler-mode chorus waves, one of the
  most important coherent electromagnetic emissions in Earth's magnetosphere.
  The simulated chorus amplitude and effective growth rate follow the predicted
  superradiant scalings, rather than exponential growth. Direct phase-space
  analysis further shows that the resonant electron trapping island remains
  centred at an approximately fixed phase during amplification. These results
  identify global phase locking as the dynamical origin of superradiant chorus
  growth and reveal a general phase-organization principle for coherent
  radiation in evolving resonant systems.
\end{abstract}

\maketitle

\section{Wave amplification and resonance in inhomogeneous systems}

\begin{figure*}
  \includegraphics[width=0.9\textwidth]{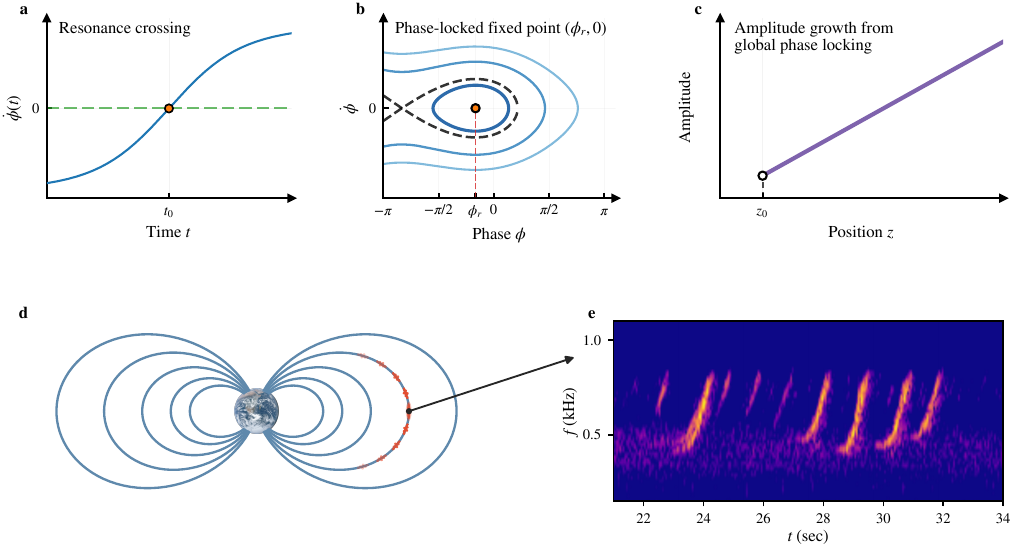}
  \caption{
    \textbf{Global phase locking and chorus emissions.}
    \textbf{a,} Resonance crossing occurs when the interaction phase angle satisfies
    $\dot{\phi}=0$ at $t=t_0$.
    \textbf{b,} Nonlinear phase locking requires $\ddot{\phi}=0$, which is
    satisfied at the stable fixed point $(\phi_r,0)$ at the centre of the trapping
    island; the dashed curve denotes the separatrix and the colored curves show
    representative trajectories.
    \textbf{c,} Global phase locking maintains the phase-space structure as the
    resonance evolves, producing linear amplitude growth with distance from an
    effective reference point $z_0$.
    \textbf{d,} Schematic of chorus wave generation and propagation along the
    terrestrial magnetic field.
    \textbf{e,} Representative frequency--time spectrogram showing discrete
    rising-tone chorus elements observed by the THEMIS satellite \cite{Angelopoulos2008} in near-Earth
    space. Color represents wave intensity; the color scale is omitted for clarity.
  }
  \label{fig1}
\end{figure*}

Coherent wave amplification often occurs in inhomogeneous systems where the
resonance condition evolves as the wave grows \cite{Kroll1981,Vomvoridis1982}. A
key variable in such interactions is the wave--particle phase angle $\phi$,
which measures the relative phase between the particle motion and the wave.
Resonance corresponds to a stationary phase, $\dot{\phi}\equiv
\mathrm{d}\phi/\mathrm{d}t=0$, allowing particles to exchange energy efficiently
with the wave (Figure~\ref{fig1}a). In the linear regime, this resonant energy
exchange can drive exponential amplification. As the wave amplitude increases,
however, resonant particles may become nonlinearly trapped by the wave field
\cite{ONeil1965}. Sustained coherent power transfer is then associated with
phase locking, $\ddot{\phi}=0$, which determines the stable resonant phase and
the associated trapping geometry \cite{Chen2021a} (Figure~\ref{fig1}b). This
hierarchy of phase conditions, however, does not by itself determine how the
wave amplitude grows once the resonance parameters evolve nonlinearly in an
inhomogeneous system. In this work, we identify a stronger phase-organization
principle, global phase locking, $\dddot{\phi}=0$, which requires the stable
resonant phase to remain nearly unchanged as the detuning evolves. This
principle leads to linear growth of field amplitude with distance, rather than
exponential growth, in the nonlinear stage (Figure~\ref{fig1}c). When the
coherent seed amplitude is negligible, the corresponding wave intensity grows
quadratically with propagation distance, as in the superradiant regime of
free-electron lasers \cite{Bonifacio1985,Giannessi2005}. Here we use the term
``superradiant'' specifically for this coherent-emission regime, rather than for
the specific many-emitter formulation of Dicke superradiance \cite{Dicke1954}.

Whistler-mode chorus waves, widely observed in space and generally believed to
arise from coherent nonlinear wave--particle interactions, provide a natural
realization of this principle. These waves are characterized by discrete,
narrowband electromagnetic emissions with rapid frequency chirping
\cite{Santolik2003} (Figure~\ref{fig1}d,e). Chorus waves are among the most
important natural emissions in Earth's magnetosphere and have been established
as central to radiation-belt electron acceleration
\cite{Horne2005b,Thorne2013b,Reeves2013} and diffuse auroral precipitation
\cite{Thorne2010}. The mechanism responsible for their frequency chirping has
been debated for many decades \cite{Helliwell1967,Nunn1974,
Vomvoridis1982,Trakhtengerts1995,Omura2008,Tao2021,Zonca2021,Zonca2022,Bonham2025}, whereas
the law governing their nonlinear amplification has received far less attention.
Where amplification has been considered, exponential wave growth is usually
assumed \cite{Vomvoridis1982,Omura2008,Tao2021,SotoChavez2012}, by analogy with linear plasma
instabilities. Whether this picture remains valid for the nonlinear excitation
of chorus, however, has not been directly tested.

Here we show that global phase locking governs the nonlinear growth of chorus
emissions. Using self-consistent kinetic simulations, we show that chorus
amplification is not exponential, contrary to the assumption often made in
previous studies. Instead, the field amplitude and effective growth rate follow
the superradiant scalings predicted by global phase locking. Direct analysis
of resonant electron phase space further shows that, within the core interaction
region where nonlinear wave amplification dominates, the stable resonant phase
remains approximately constant as the wave amplitude increases. Together, these
results identify global phase locking as the mechanism underlying superradiant
chorus growth and suggest a broader phase-organization principle for coherent
amplification in evolving resonant systems.

\section{Global phase locking}

We consider a coherent wave--particle interaction in a weakly inhomogeneous
system, with the dominant inhomogeneity along the coordinate $z$. The nonlinear
phase dynamics near resonance can be written in the generic driven-pendulum form
\cite{Neishtadt2013,Artemyev2023}
\begin{align}
  \label{eq:ddot_phi}
  \ddot{\phi}=A\sin\phi-\tau(z).
\end{align}
Here $A$ measures the nonlinear trapping strength and is proportional to the
wave amplitude, while $\tau$ represents an external detuning term, namely the
slow change in the resonance condition caused by magnetic-field inhomogeneity,
frequency chirping, or other slowly varying system parameters. By weak
inhomogeneity, we mean that $\tau$ varies on a spatial scale much longer than
the nonlinear amplification scale. A derivation of Equation~\eqref{eq:ddot_phi}
from a general Hamiltonian formulation is given in Section
\ref{sec:forced_pendulum_hamiltonian} of the Appendix. When $|\tau|<A$,
Equation~\eqref{eq:ddot_phi} admits nonlinear phase trapping. The stable phase
of the trapping island, $\phi_r$, is determined by the phase-locking condition
$\ddot{\phi}=0$, or equivalently $\sin\phi_r=\tau(z)/A(z)$. This phase-locking
condition is local because it determines $\phi_r$ at a given position or time.
The value of $\phi_r$ controls the trapping geometry (Figure~\ref{fig1}b) and is
related to the maximization of wave--particle power transfer \cite{Vomvoridis1982}.

In an inhomogeneous system, both the external detuning $\tau$ and the amplitude
$A$ evolve as the wave propagates. If the wave--particle power transfer
associated with phase locking is to remain near its maximum during propagation,
the stable phase $\phi_r$ must stay nearly unchanged as the detuning and wave
amplitude evolve. We define this persistence of the stable resonant phase as
global phase locking. Since $\sin\phi_r=\tau(z)/A(z)$, maintaining an
approximately constant $\phi_r$ requires the wave amplitude to track the
evolving detuning, so that $\tau/A$ remains approximately constant. In dynamical
terms, this corresponds to maintaining the phase-locking condition at the next
order, $\dddot{\phi}=0$. Over the core interaction region, the slowly varying
detuning can be expanded about a reference point $z_{0}$ as $\tau(z)\simeq
\tau(z_0)+\tau'(z_0)(z-z_0)$. Thus $A(z)$ follows the leading spatial variation
of the detuning,
\begin{align}
  \label{eq:linear_A}
  A(z)\simeq A(z_{0})+C(z-z_0).
\end{align}
Here $A(z_0)\simeq \tau(z_0)/\sin\phi_r$ and $C\simeq \tau'(z_0)/\sin\phi_r$.
Amplitude tracking, $A\propto\tau$, is therefore the general consequence of
global phase locking, while the linear-amplitude law in
Equation~\eqref{eq:linear_A} is its leading-order form for a slowly varying
detuning.

The resulting linear-amplitude law is a central prediction of global phase locking
(Figure~\ref{fig1}c). It states that coherent amplification in this regime is
not exponential, but linear in distance from a reference point. In the chorus
simulations analyzed below, this reference point corresponds to the upstream
onset of coherent nonlinear amplification, where the wave amplitude is close to
the noise floor, so that $A(z_0)\simeq 0$. The effective growth rate then
becomes
\begin{align}
  \label{sr_gamma}
  \gamma
  \equiv
  \frac{1}{A}\frac{\mathrm{d}A}{\mathrm{d}t}
  \simeq
  \frac{v_g}{z-z_0},
\end{align}
where $v_g$ is the wave group velocity. The wave intensity therefore satisfies
\begin{align}
  I(z)\propto A^2(z)\propto (z-z_0)^2 .
\end{align}
This quadratic intensity growth is the characteristic scaling of superradiant
emission \cite{Bonifacio1985}. The same linear amplitude evolution of chorus
waves can be derived from the TaRA model by relating two different expressions
for the frequency chirping rate (Section~\ref{linear_amplitude_from_tara} of the
Appendix). In that formulation, however, the broader significance of this result
was not recognized. The derivation given here shows that the linear amplitude
law follows from the more general principle of global phase locking.

\section{Superradiant growth in chorus waves}

We test the global phase-locking prediction using self-consistent kinetic
simulations of whistler-mode chorus waves. Chorus is particularly well suited
to this analysis because its coherent nonlinear amplification remains poorly
understood and because the combined effects of magnetic-field inhomogeneity and
frequency chirping cause the resonance condition to evolve continuously during
wave growth.

\begin{figure*}
  \includegraphics[width=0.9\textwidth]{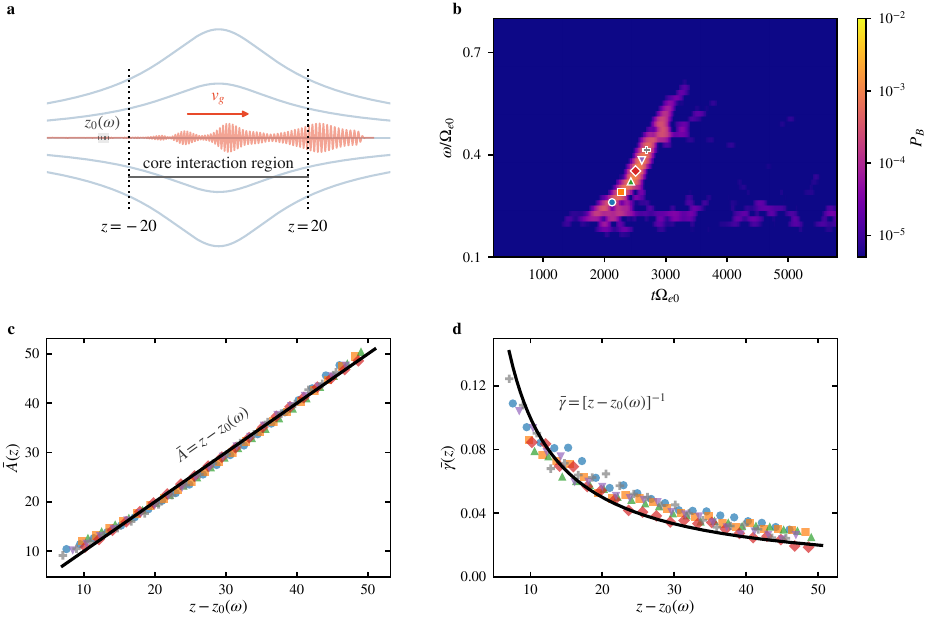} 
  \caption{ 
    \textbf{Superradiant
    growth of chorus waves in a self-consistent kinetic simulation.} \textbf{a,}
    Schematic of the simulation geometry. The background magnetic field has a
    mirror-like inhomogeneity along the $z$ direction, and the chorus wave
    packet propagates from the upstream source region toward the magnetic
    equator with group velocity $v_g$. The effective source location for each
    frequency is denoted by $z_0(\omega)$. The dotted vertical lines delimit the
    core interaction region, $-20 \leq z \leq 20$, used for the amplitude and
    growth-rate analyses in \textbf{c} and \textbf{d}. \textbf{b,} Wave
    magnetic-field spectrogram measured at the magnetic equator, showing a
    rising-tone chorus element. The colored symbols indicate the selected
    frequencies along the chirping element used for the spatial growth analysis.
    \textbf{c,} Normalized wave amplitude $\bar{A}(z)$ as a function of
    $z-z_0(\omega)$. Results at the selected frequencies collapse onto the
    predicted linear scaling, $\bar{A}=z-z_0(\omega)$, with a relative
    root-mean-square (RMS) deviation of $2.7\%$. \textbf{d,} Normalized growth
    rate $\bar{\gamma}=\gamma/v_g$ versus $z-z_0(\omega)$.  The simulation
    results follow the inverse-distance scaling
    $\bar{\gamma}=[z-z_0(\omega)]^{-1}$ predicted for superradiant growth, with
    a relative RMS deviation of $15.8\%$. The relative RMS deviation is defined
  in the Appendix.  
} 
\label{fig2}
\end{figure*}

The simulation uses a one-dimensional spatial domain along the background
magnetic field and three-dimensional particle velocities \cite{Tao2014b}. The
background magnetic field has a mirror-like profile that approximates the
dipole-field strength variation near the magnetic equator, $B(z)=B_0(1+\xi
z^2)$, where $z$ is the distance along the field normalized by $c/\Omega_{e0}$, 
$c$ is the speed of light in vacuum, $\Omega_{e0}$ is the equatorial electron
gyrofrequency, $\xi$ characterizes the magnetic-field inhomogeneity, and $z=0$
corresponds to the equator. As illustrated in Figure~\ref{fig2}a, waves
originate in the upstream region ($z<0$) and are amplified as they propagate
downstream. We denote the frequency-dependent effective onset location of
coherent nonlinear amplification in the upstream region by $z_0(\omega)$.
Because the wave amplitude remains close to the noise level near this location,
$z_0(\omega)$ cannot be exactly identified. As the wave packet propagates,
however, it undergoes substantial amplification and develops into a clear
rising-tone chorus element, as shown in the spectrogram in Figure~\ref{fig2}b.
We restrict the quantitative analysis to the core interaction region, $-20\leq
z\leq 20$. Farther upstream, the wave signal is too weak for reliable
measurement, whereas farther downstream, boundary effects become increasingly
important.

Within this region, we determine the spatial amplification law by analysing the
wave amplitude and growth rate at six approximately equally spaced frequencies
along the chorus element (Figure~\ref{fig2}b). Because the wave amplitude and
effective growth rate depend on frequency, we normalize both quantities to
compare their predicted spatial scalings across all six frequencies. Based on
Equation~\eqref{eq:linear_A}, the normalized amplitude is
\begin{align}
  \bar{A}(z) \equiv \frac{A(z)-A(z_0)}{C} = z-z_0(\omega),
\end{align}
while, because $A(z_0)$ is close to the numerical noise level,
Equation~\eqref{sr_gamma} gives
\begin{align}
  \bar{\gamma} \equiv \frac{\gamma}{v_g} \simeq \frac{1}{z-z_0(\omega)}.
\end{align}
For each selected frequency, we first fit the measured normalized growth rate to
the inverse-distance scaling to determine $z_0(\omega)$. Using this value, we
then fit the corresponding wave amplitude to $A(z)=A(z_0)+C[z-z_0(\omega)]$ to
determine $A(z_0)$ and $C$, from which $\bar{A}$ is constructed. The comparison
therefore tests both the linear amplitude scaling and the inverse-distance
growth-rate scaling predicted by global phase locking.

Figures~\ref{fig2}c and~\ref{fig2}d show that the normalized amplitudes at all
six frequencies collapse closely onto the predicted linear scaling, while the
normalized growth rates follow the inverse-distance dependence predicted by
global phase locking. This agreement is obtained across multiple frequencies
within the chirping element. The effective growth rate varies by approximately a
factor of four across this region, ruling out an interpretation of the observed
linear amplitude scaling as a local approximation to exponential growth with a
constant growth rate. Over the analyzed core interaction region, chorus
amplification is therefore well described by linear field-amplitude growth with
distance, rather than exponential growth, implying quadratic growth of wave
intensity characteristic of superradiant emission.

\section{Phase-space evidence for global phase locking}

The amplitude and growth-rate scalings demonstrate that chorus waves exhibit
superradiant amplification. We next test the proposed mechanism directly by
examining the nonlinear phase-space structure of resonant electrons. If
superradiant growth is caused by global phase locking, then the phase-space
structure associated with nonlinear trapping should remain organized around an
approximately fixed resonant phase.

\begin{figure*}
  \includegraphics[width=0.9\textwidth]{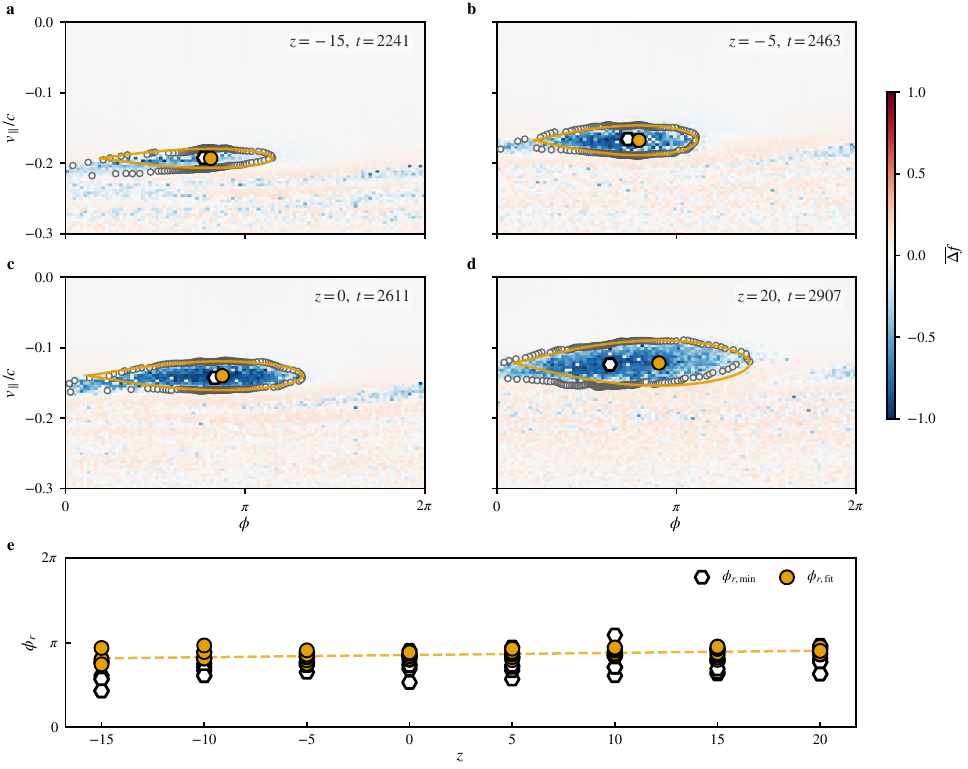}  
  \caption{ 
    \textbf{Phase-space evidence for global phase locking.} \textbf{a--d,}
    Electron phase-space distributions in the $(\phi,v_{\parallel})$ plane at
    four representative positions and times. The color scale shows the
    normalized phase-space density perturbation $\overline{\Delta
    f}=(f-f_0)/\max|f-f_0|$, where $f$ is the instantaneous electron
    distribution and $f_0$ is the initial unperturbed distribution. The blue
    regions show the phase-space depletion associated with nonlinear trapping of
    resonant electrons. White circles mark the estimated island boundary, and
    the golden line shows the fitted boundary. The hexagons and filled golden
    circles indicate $\phi_r$ determined from the minimum of $\overline{\Delta
    f}$ and from the boundary fit, respectively. The fitting uncertainties in
    $\phi_{r,\mathrm{fit}}$, estimated from the local covariance matrix, are of
    order $10^{-3}\,\mathrm{rad}$ and are smaller than the plotted symbols.
    \textbf{e,} Resonant phase $\phi_r$ at different positions and times. The
    golden dashed line shows a linear fit to the fitted resonant phases,
    $\phi_r=(2.68\pm0.02)+(0.0081\pm0.0016)z$, where the uncertainties are
    standard errors from the linear regression. The fitted phase remains close
    to $\phi_r\simeq2.68$ throughout the core interaction region, showing that
    phase locking persists as the wave amplifies. 
  }     
\label{fig3}
\end{figure*}

We construct electron phase-space distributions in the coordinates
$(\phi,v_\parallel)$ at multiple locations across the core interaction region
and at multiple times during the formation of the chorus element (see
Section~\ref{kinetic_simulation} of the Appendix),
where $v_\parallel$ is the electron velocity parallel to the background magnetic
field. Near resonance, $\dot{\phi}\simeq k(v_\parallel-v_r)$, where $k$ is the
parallel wave number and $v_r$ is the resonant parallel velocity. Plotting
$v_\parallel$ is therefore equivalent, up to a shift and scale factor, to
plotting $\dot{\phi}$, while avoiding the need to determine $v_r$ separately at
each time and location. In these coordinates, nonlinear trapping appears as a
phase-space hole associated with the trapping island, similar to the structure
shown schematically in Figure~\ref{fig1}b. The distributions become noisier
farther upstream, where the wave amplitude is weaker and the phase-space hole is
less developed. As a result, not all sampled locations, including the upstream
edge of the core interaction region, yield a clearly identifiable trapping
boundary. Four representative distributions are shown in Figures~\ref{fig3}a--d.

We determine the resonant phase $\phi_r$ using two complementary methods. The
first identifies the location of the minimum phase-space density inside the
hole, giving the estimate $\phi_{r,\min}$. This provides a direct measure of the
hole centre but is sensitive to particle noise. The second estimates the
boundary of the phase-space hole from the particle distribution and fits it
with the nonlinear trapping separatrix predicted by the driven-pendulum
equation, giving the estimate $\phi_{r,\mathrm{fit}}$. This boundary-fitting
method provides a more robust estimate of the hole centre, particularly when
the local density minimum is noisy. Details of the boundary extraction and both
estimation procedures are provided in Section~\ref{kinetic_simulation} of the
Appendix.

The resulting resonant phases are shown in Figure~\ref{fig3}e. The two estimates
agree most closely near the equator and differ more toward the edges of the core
interaction region, owing to the weaker wave amplitude upstream and the more
distorted hole structure farther downstream. Most importantly,
$\phi_{r,\mathrm{fit}}$ remains approximately constant over the broad interval
$-15\lesssim z\lesssim 20$, with $\phi_r\simeq 2.68$ and $\sin\phi_r\simeq 0.43$,
close to the value associated with maximum wave--particle power transfer
\cite{Vomvoridis1982,Omura2008,Tao2021,Zonca2022}. This
nearly constant resonant phase is the phase-space signature of global phase
locking. Although the detuning evolves as the wave propagates and its frequency
chirps, the phase-space hole remains centred at approximately the same phase
while the wave amplitude increases. The amplitude growth therefore compensates
the evolving detuning and preserves the trapping geometry associated with
efficient wave--particle power transfer. The simulation thus reveals both the
superradiant growth law and the phase-space organization responsible for it.

\section{Implications for coherent radiation}

The superradiant interpretation changes the physical picture of nonlinear chorus
amplification. In the core nonlinear interaction region, chorus growth is not
exponential, contrary to assumptions often made in previous studies. Instead,
it is governed by the coupled evolution of the resonant phase and the wave
amplitude. To maintain near-optimal wave--particle power transfer, global
phase locking requires the resonant phase to remain nearly unchanged as the
detuning evolves, forcing the wave amplitude to adjust accordingly and producing
superradiant growth. This interpretation also clarifies the underlying
principle behind the linear amplitude law obtained from the TaRA model.

Although demonstrated here for chorus waves, global phase locking is not
specific to magnetospheric plasma physics. Its broad applicability comes from
the ubiquity of pendulum-like nonlinear phase-space structures in resonant
wave--particle interactions, as made explicit by the Hamiltonian derivation
shown in the Appendix. Global phase locking therefore provides a dynamical
link between microscopic phase-space organization and macroscopic coherent-growth
laws, showing how superradiant amplification can arise in an evolving resonant
system. Whether the same mechanism operates in other slowly evolving resonant
systems, including free-electron lasers, is a natural subject for future
investigation.

\section{Appendix}

\subsection{Hamiltonian derivation of the driven-pendulum equation}
\label{sec:forced_pendulum_hamiltonian}

Near resonance, the dynamics of a broad class of slowly varying Hamiltonian
systems can be reduced to a pendulum with an effective torque
\cite{Lichtenberg1983,Neishtadt2013,Artemyev2018,Artemyev2023}.
We write the reduced resonant Hamiltonian as
\begin{align}
  H(I,\phi,z,P_z,t)
  =
  H_0(I;z,P_z,t)
  +
  V(I;z,P_z,t)\cos\phi,
  \label{eq:reduced_resonant_hamiltonian}
\end{align}
where $(I,\phi)$ is the resonant action--angle pair and $(z,P_z)$ denote
slow phase-space variables. The instantaneous resonant action $I_r(z,P_z,t)$ is defined by
\begin{align}
  \left.
  \frac{\partial H_0}{\partial I}
  \right|_{I=I_r}
  =0.
\end{align}
Expanding about $I=I_r$ and retaining the leading resonant terms gives
\begin{align}
  H
  \simeq
  H_0(I_r)
  +
  \frac{1}{2}g(\Delta I)^2
  +
  V\cos\phi,
\end{align}
where $\Delta I=I-I_r$,
$g=\left.\partial^2H_0/\partial I^2\right|_{I_r}$, and $V$ is evaluated at
resonance. Hamilton's equations then give
\begin{align}
  \dot\phi\simeq g\Delta I,
  \qquad
  \dot I=V\sin\phi.
\end{align}
Taking another time derivative and neglecting the slow variation of $g$ over
a trapping period yields
\begin{align}
  \ddot\phi
  \simeq
  gV\sin\phi-g\dot I_r,
  \label{eq:forced_pendulum_general}
\end{align}
where
\begin{align}
  \dot I_r
  =
  \frac{\partial I_r}{\partial t}
  +
  \dot z\frac{\partial I_r}{\partial z}
  +
  \dot P_z\frac{\partial I_r}{\partial P_z}.
\end{align}
The explicit temporal term includes, for example, resonance drift associated
with frequency chirping, while the remaining terms describe drift through the
inhomogeneous background. Comparing Equation~\eqref{eq:forced_pendulum_general}
with Equation~\eqref{eq:ddot_phi} gives
\begin{align}
  A=gV, \quad \tau=g\dot I_r.
\end{align}
Along the relevant resonant trajectory, the slow variables $P_z$ and $t$ may
be parameterized by $z$, giving the effective functions $A(z)$ and $\tau(z)$
used in the main text.

\subsection{Derivation of the linear amplitude law from the TaRA model}
\label{linear_amplitude_from_tara}

The TaRA model \cite{Tao2021} predicts two expressions for the frequency chirping rate of
magnetospheric chorus. Assuming that the cold-plasma density is constant along
the field line, the first is determined by the magnetic-field gradient \cite{Helliwell1967} at the
upstream source location $z_0$:
\begin{align}
\frac{\partial\omega}{\partial t}
\approx
-\left.
\left(1-\frac{v_r}{v_g}\right)^{-2}
\left(
\frac{k v_{\perp}^{2}}{2\Omega_e}
-\frac{3v_r}{2}
\right)
\frac{\partial\Omega_e}{\partial z}
\right|_{z_0}.
\end{align}
Here, $v_r$ is the resonant parallel velocity, $k$ is the wave number, and
$v_\perp$ is the characteristic perpendicular velocity of the resonant
electrons. The second expression gives the nonlinear chirping rate at a
position $z$ within the core interaction region \cite{Vomvoridis1982}:
\begin{align}
\frac{\partial\omega}{\partial t}
\approx
\left(1-\frac{v_r}{v_g}\right)^{-2}
\left[
\frac{1}{2}\omega_{\mathrm{tr}}^{2}
-
\left(
\frac{k v_{\perp}^{2}}{2\Omega_e}
-\frac{3v_r}{2}
\right)
\frac{\partial\Omega_e}{\partial z}
\right].
\end{align}
Since the chorus source region is typically confined to within approximately
$3^{\circ}$ of the equator \cite{Santolik2003}, the spatial variation of the frequency chirping rate
is small. The two chirping rates are therefore approximately equal, giving
\begin{align}
  \frac{\delta B}{B_{0}} \approx 2 \left.\frac{\xi}{k} \left(\frac{k v_{\perp}}{\Omega_{e}} 
    - 3 \frac{v_{r}}{v_{\perp}}\right)\right|_{z_{0}} (z-z_{0}).
\end{align}
The coefficient varies slowly and may be evaluated at any point in the
interaction region, such as the equator, without affecting the leading-order
result. We evaluate it at $z_0$, consistent with Equation~\eqref{eq:linear_A}.
The TaRA model therefore predicts that the chorus wave amplitude increases
linearly with the distance from $z_0$, a result previously used by
\cite{Wu2023} to estimate chorus amplitudes near the magnetic equator. 
Direct application of the global phase-locking condition
to the chorus pendulum equation \cite{Tao2021} yields the same result.

\subsection{Kinetic simulation and analysis of chorus waves}
\label{kinetic_simulation}

The chorus simulation data analyzed in this work were generated before the
global phase-locking framework was formulated, and no
simulation parameters were changed for the present study \cite{Tao2021}. The
results therefore represent a new analysis of an existing self-consistent
simulation rather than a simulation tuned to reproduce the predicted scalings.
The simulation was performed with DAWN, a one-dimensional particle simulation
code developed for parallel-propagating chorus waves
\cite{Tao2014b,Tao2017b,Tao2021}. The background magnetic field is 
$B(z)=B_0(1+\xi z^2)$, with length and time normalized by $c/\Omega_{e0}$
and $\Omega_{e0}^{-1}$, respectively. In these normalized units,
$\xi=2.155\times10^{-5}$, corresponding to the dipole magnetic field of a planet
with a radius one-sixteenth that of Earth's, evaluated on the $L=4$ field line.
The reduced planetary scale is adopted to lower the computational cost. The
computational domain has a total length of approximately $328$, placing the
simulation boundaries far from the core interaction region and therefore
minimizing boundary effects. Full details of the numerical model, simulation
parameters, and configuration are given in Ref.~\cite{Tao2021}.

The six representative frequencies analyzed in this study, as indicated in
Figure~\ref{fig2}b, correspond to Fourier-frequency bins at
$\omega/\Omega_{e0}=0.261$, $0.291$, $0.322$, $0.353$, $0.383$, and $0.414$. For
the phase-space analysis in Figure~\ref{fig3}e, we examine ten
particle-distribution snapshots saved at approximately equal time intervals
between $t\approx2241$ and $2907$, covering the generation and nonlinear
evolution of the chorus element. At each selected time, the distributions are
analyzed from $z=-20$ to $20$ with $\Delta z=5$. At $z=-20$, no clear
phase-space depletion can be identified because of the weak wave signal and
simulation noise. To quantify the agreement between the simulation results and
the theoretical scalings in Figures~\ref{fig2}c and \ref{fig2}d, we use the
relative RMS
deviation,
\begin{align}
  \epsilon_{\mathrm{RMS}}
  =
  \left[
    \frac{
      \sum_i \left(y_i-y_{\mathrm{th},i}\right)^2
    }{
      \sum_i y_{\mathrm{th},i}^2
    }
  \right]^{1/2},
\end{align}
where $y_i$ denotes the simulation result and $y_{\mathrm{th},i}$ the
corresponding theoretical prediction.

The centre and boundary of the phase-space depletion are determined directly
from the phase-space density perturbation, $\Delta f$, in $(\phi,v_{\parallel})$
coordinates, constructed using $100\times105$ bins. To reduce particle noise,
the two-dimensional distribution is first smoothed with a Gaussian kernel of one
grid cell in each direction. A $5\times5$ local average is then applied, and the
minimum of the averaged distribution defines the depletion-core centre,
$(\phi_{r,\min},v_{r,\min})$.    Starting from this point, 240 uniformly spaced
radial lines are sampled through the smoothed distribution. Along each ray, the
boundary is identified as the first outward crossing of the threshold $\Delta
f_{\mathrm{th}}=0.2 \Delta f_{\min}$, where $\Delta f_{\min}<0$ is the locally
averaged value at the depletion centre. Linear interpolation between adjacent
samples is used to determine the crossing position.

The extracted boundary is subsequently fitted to the separatrix of the
driven-pendulum model derived above. The fit parameters are the stable-centre
phase $\phi_{r,\mathrm{fit}}$, the corresponding parallel velocity
$v_{r,\mathrm{fit}}$, and the trapping width $\Delta v_{\mathrm{tr}}$. For each
selected boundary point $(\phi_i,v_i)$, the residual is defined as
\begin{equation}
  \epsilon_i \equiv
  \Delta v_{\mathrm{tr}}
  \sqrt{
    2\left[
      \cos\phi_s-\cos\phi_i
      -R(\phi_i-\phi_s)
    \right]
  }
  -\left|v_i-v_{r,\mathrm{fit}}\right|.
\end{equation}
Here, $R$ and $\phi_s$ are given by
\begin{align}
  R&=\sin\phi_{r,\mathrm{fit}},
  \qquad
  \phi_s=\pi-\phi_{r,\mathrm{fit}}.
\end{align}
The fitting parameters are determined by jointly minimizing a robust loss
function constructed from the residuals over the estimated boundary points.
Boundary points within the phase interval $1\leq\phi\leq5$ are used by default,
with limited case-specific adjustments to the fitting interval and weighting
when the extracted boundary is incomplete or asymmetric.

%\thebibliography{unsrt}
% \bibliography{refs}

\begin{thebibliography}{29}%
\makeatletter
\providecommand \@ifxundefined [1]{%
 \@ifx{#1\undefined}
}%
\providecommand \@ifnum [1]{%
 \ifnum #1\expandafter \@firstoftwo
 \else \expandafter \@secondoftwo
 \fi
}%
\providecommand \@ifx [1]{%
 \ifx #1\expandafter \@firstoftwo
 \else \expandafter \@secondoftwo
 \fi
}%
\providecommand \natexlab [1]{#1}%
\providecommand \enquote  [1]{``#1''}%
\providecommand \bibnamefont  [1]{#1}%
\providecommand \bibfnamefont [1]{#1}%
\providecommand \citenamefont [1]{#1}%
\providecommand \href@noop [0]{\@secondoftwo}%
\providecommand \href [0]{\begingroup \@sanitize@url \@href}%
\providecommand \@href[1]{\@@startlink{#1}\@@href}%
\providecommand \@@href[1]{\endgroup#1\@@endlink}%
\providecommand \@sanitize@url [0]{\catcode `\\12\catcode `\$12\catcode `\&12\catcode `\#12\catcode `\^12\catcode `\_12\catcode `\%12\relax}%
\providecommand \@@startlink[1]{}%
\providecommand \@@endlink[0]{}%
\providecommand \url  [0]{\begingroup\@sanitize@url \@url }%
\providecommand \@url [1]{\endgroup\@href {#1}{\urlprefix }}%
\providecommand \urlprefix  [0]{URL }%
\providecommand \Eprint [0]{\href }%
\providecommand \doibase [0]{https://doi.org/}%
\providecommand \selectlanguage [0]{\@gobble}%
\providecommand \bibinfo  [0]{\@secondoftwo}%
\providecommand \bibfield  [0]{\@secondoftwo}%
\providecommand \translation [1]{[#1]}%
\providecommand \BibitemOpen [0]{}%
\providecommand \bibitemStop [0]{}%
\providecommand \bibitemNoStop [0]{.\EOS\space}%
\providecommand \EOS [0]{\spacefactor3000\relax}%
\providecommand \BibitemShut  [1]{\csname bibitem#1\endcsname}%
\let\auto@bib@innerbib\@empty
%</preamble>
\bibitem [{\citenamefont {Kroll}\ \emph {et~al.}(1981)\citenamefont {Kroll}, \citenamefont {Morton},\ and\ \citenamefont {Rosenbluth}}]{Kroll1981}%
  \BibitemOpen
  \bibfield  {author} {\bibinfo {author} {\bibfnamefont {N.}~\bibnamefont {Kroll}}, \bibinfo {author} {\bibfnamefont {P.}~\bibnamefont {Morton}},\ and\ \bibinfo {author} {\bibfnamefont {M.}~\bibnamefont {Rosenbluth}},\ }\bibfield  {title} {\bibinfo {title} {Free-electron lasers with variable parameter wigglers},\ }\href {https://doi.org/10.1109/JQE.1981.1071285} {\bibfield  {journal} {\bibinfo  {journal} {IEEE Journal of Quantum Electronics}\ }\textbf {\bibinfo {volume} {17}},\ \bibinfo {pages} {1436} (\bibinfo {year} {1981})}\BibitemShut {NoStop}%
\bibitem [{\citenamefont {Vomvoridis}\ \emph {et~al.}(1982)\citenamefont {Vomvoridis}, \citenamefont {Crystal},\ and\ \citenamefont {Denavit}}]{Vomvoridis1982}%
  \BibitemOpen
  \bibfield  {author} {\bibinfo {author} {\bibfnamefont {J.~L.}\ \bibnamefont {Vomvoridis}}, \bibinfo {author} {\bibfnamefont {T.~L.}\ \bibnamefont {Crystal}},\ and\ \bibinfo {author} {\bibfnamefont {J.}~\bibnamefont {Denavit}},\ }\bibfield  {title} {\bibinfo {title} {Theory and computer simulations of magnetospheric very low frequency emissions},\ }\href {https://doi.org/10.1029/JA087iA03p01473} {\bibfield  {journal} {\bibinfo  {journal} {J.~Geophys. Res.}\ }\textbf {\bibinfo {volume} {87}},\ \bibinfo {pages} {1473} (\bibinfo {year} {1982})}\BibitemShut {NoStop}%
\bibitem [{\citenamefont {O'Neil}(1965)}]{ONeil1965}%
  \BibitemOpen
  \bibfield  {author} {\bibinfo {author} {\bibfnamefont {T.}~\bibnamefont {O'Neil}},\ }\bibfield  {title} {\bibinfo {title} {Collisionless damping of nonlinear plasma oscillations},\ }\href {https://doi.org/10.1063/1.1761193} {\bibfield  {journal} {\bibinfo  {journal} {Phys. Fluids}\ }\textbf {\bibinfo {volume} {8}},\ \bibinfo {pages} {2255} (\bibinfo {year} {1965})}\BibitemShut {NoStop}%
\bibitem [{\citenamefont {{Chen}}\ \emph {et~al.}(2021)\citenamefont {{Chen}}, \citenamefont {{Zonca}},\ and\ \citenamefont {{Lin}}}]{Chen2021a}%
  \BibitemOpen
  \bibfield  {author} {\bibinfo {author} {\bibfnamefont {L.}~\bibnamefont {{Chen}}}, \bibinfo {author} {\bibfnamefont {F.}~\bibnamefont {{Zonca}}},\ and\ \bibinfo {author} {\bibfnamefont {Y.}~\bibnamefont {{Lin}}},\ }\bibfield  {title} {\bibinfo {title} {{Physics of kinetic Alfv{\'e}n waves: a gyrokinetic theory approach}},\ }\href {https://doi.org/10.1007/s41614-020-00049-3} {\bibfield  {journal} {\bibinfo  {journal} {Reviews of Modern Plasma Physics}\ }\textbf {\bibinfo {volume} {5}},\ \bibinfo {eid} {1} (\bibinfo {year} {2021})}\BibitemShut {NoStop}%
\bibitem [{\citenamefont {Bonifacio}\ and\ \citenamefont {Casagrande}(1985)}]{Bonifacio1985}%
  \BibitemOpen
  \bibfield  {author} {\bibinfo {author} {\bibfnamefont {R.}~\bibnamefont {Bonifacio}}\ and\ \bibinfo {author} {\bibfnamefont {F.}~\bibnamefont {Casagrande}},\ }\bibfield  {title} {\bibinfo {title} {The superradiant regime of a free electron laser},\ }\href {https://doi.org/https://doi.org/10.1016/0168-9002(85)90695-3} {\bibfield  {journal} {\bibinfo  {journal} {Nuclear Instruments and Methods in Physics Research Section A: Accelerators, Spectrometers, Detectors and Associated Equipment}\ }\textbf {\bibinfo {volume} {239}},\ \bibinfo {pages} {36} (\bibinfo {year} {1985})}\BibitemShut {NoStop}%
\bibitem [{\citenamefont {Giannessi}\ \emph {et~al.}(2005)\citenamefont {Giannessi}, \citenamefont {Musumeci},\ and\ \citenamefont {Spampinati}}]{Giannessi2005}%
  \BibitemOpen
  \bibfield  {author} {\bibinfo {author} {\bibfnamefont {L.}~\bibnamefont {Giannessi}}, \bibinfo {author} {\bibfnamefont {P.}~\bibnamefont {Musumeci}},\ and\ \bibinfo {author} {\bibfnamefont {S.}~\bibnamefont {Spampinati}},\ }\bibfield  {title} {\bibinfo {title} {Nonlinear pulse evolution in seeded free-electron laser amplifiers and in free-electron laser cascades},\ }\href {https://doi.org/10.1063/1.2010624} {\bibfield  {journal} {\bibinfo  {journal} {Journal of Applied Physics}\ }\textbf {\bibinfo {volume} {98}},\ \bibinfo {pages} {043110} (\bibinfo {year} {2005})}\BibitemShut {NoStop}%
\bibitem [{\citenamefont {Dicke}(1954)}]{Dicke1954}%
  \BibitemOpen
  \bibfield  {author} {\bibinfo {author} {\bibfnamefont {R.~H.}\ \bibnamefont {Dicke}},\ }\bibfield  {title} {\bibinfo {title} {Coherence in spontaneous radiation processes},\ }\href {https://doi.org/10.1103/PhysRev.93.99} {\bibfield  {journal} {\bibinfo  {journal} {Phys. Rev.}\ }\textbf {\bibinfo {volume} {93}},\ \bibinfo {pages} {99} (\bibinfo {year} {1954})}\BibitemShut {NoStop}%
\bibitem [{\citenamefont {Santol\'{i}k}\ \emph {et~al.}(2003)\citenamefont {Santol\'{i}k}, \citenamefont {Gurnett}, \citenamefont {Pickett}, \citenamefont {Parrot},\ and\ \citenamefont {Cornilleau-Wehrlin}}]{Santolik2003}%
  \BibitemOpen
  \bibfield  {author} {\bibinfo {author} {\bibfnamefont {O.}~\bibnamefont {Santol\'{i}k}}, \bibinfo {author} {\bibfnamefont {D.~A.}\ \bibnamefont {Gurnett}}, \bibinfo {author} {\bibfnamefont {J.~S.}\ \bibnamefont {Pickett}}, \bibinfo {author} {\bibfnamefont {M.}~\bibnamefont {Parrot}},\ and\ \bibinfo {author} {\bibfnamefont {N.}~\bibnamefont {Cornilleau-Wehrlin}},\ }\bibfield  {title} {\bibinfo {title} {Spatio-temporal structure of storm-time chorus},\ }\href {https://doi.org/10.1029/2002JA009791} {\bibfield  {journal} {\bibinfo  {journal} {J.~Geophys. Res.}\ }\textbf {\bibinfo {volume} {108}},\ \bibinfo {pages} {1278} (\bibinfo {year} {2003})}\BibitemShut {NoStop}%
\bibitem [{\citenamefont {Horne}\ \emph {et~al.}(2005)\citenamefont {Horne}, \citenamefont {Thorne}, \citenamefont {Shprits}, \citenamefont {Meredith}, \citenamefont {Glauert}, \citenamefont {Smith}, \citenamefont {Kanekal}, \citenamefont {Baker}, \citenamefont {Engebretson}, \citenamefont {Posch}, \citenamefont {Spasojevic}, \citenamefont {Inan}, \citenamefont {Pickett},\ and\ \citenamefont {Decreau}}]{Horne2005b}%
  \BibitemOpen
  \bibfield  {author} {\bibinfo {author} {\bibfnamefont {R.~B.}\ \bibnamefont {Horne}}, \bibinfo {author} {\bibfnamefont {R.~M.}\ \bibnamefont {Thorne}}, \bibinfo {author} {\bibfnamefont {Y.~Y.}\ \bibnamefont {Shprits}}, \bibinfo {author} {\bibfnamefont {N.~P.}\ \bibnamefont {Meredith}}, \bibinfo {author} {\bibfnamefont {S.~A.}\ \bibnamefont {Glauert}}, \bibinfo {author} {\bibfnamefont {A.~J.}\ \bibnamefont {Smith}}, \bibinfo {author} {\bibfnamefont {S.~G.}\ \bibnamefont {Kanekal}}, \bibinfo {author} {\bibfnamefont {D.~N.}\ \bibnamefont {Baker}}, \bibinfo {author} {\bibfnamefont {M.~J.}\ \bibnamefont {Engebretson}}, \bibinfo {author} {\bibfnamefont {J.~L.}\ \bibnamefont {Posch}}, \bibinfo {author} {\bibfnamefont {M.}~\bibnamefont {Spasojevic}}, \bibinfo {author} {\bibfnamefont {U.~S.}\ \bibnamefont {Inan}}, \bibinfo {author} {\bibfnamefont {J.~S.}\ \bibnamefont {Pickett}},\ and\ \bibinfo {author} {\bibfnamefont {P.~M.~E.}\ \bibnamefont {Decreau}},\ }\bibfield  {title} {\bibinfo {title} {Wave acceleration of electrons in the {V}an {A}llen radiation belts},\ }\href {https://doi.org/10.1038/nature03939} {\bibfield  {journal} {\bibinfo  {journal} {Nature}\ }\textbf {\bibinfo {volume} {437}},\ \bibinfo {pages} {227} (\bibinfo {year} {2005})}\BibitemShut {NoStop}%
\bibitem [{\citenamefont {Thorne}\ \emph {et~al.}(2013)\citenamefont {Thorne}, \citenamefont {Li}, \citenamefont {Ni}, \citenamefont {Ma}, \citenamefont {Bortnik}, \citenamefont {Chen}, \citenamefont {Baker}, \citenamefont {Spence}, \citenamefont {Reeves}, \citenamefont {Henderson}, \citenamefont {Kletzing}, \citenamefont {Kurth}, \citenamefont {Hospodarsky}, \citenamefont {Blake}, \citenamefont {Fennell}, \citenamefont {Claudepierre},\ and\ \citenamefont {Kanekal}}]{Thorne2013b}%
  \BibitemOpen
  \bibfield  {author} {\bibinfo {author} {\bibfnamefont {R.~M.}\ \bibnamefont {Thorne}}, \bibinfo {author} {\bibfnamefont {W.}~\bibnamefont {Li}}, \bibinfo {author} {\bibfnamefont {B.}~\bibnamefont {Ni}}, \bibinfo {author} {\bibfnamefont {Q.}~\bibnamefont {Ma}}, \bibinfo {author} {\bibfnamefont {J.}~\bibnamefont {Bortnik}}, \bibinfo {author} {\bibfnamefont {L.}~\bibnamefont {Chen}}, \bibinfo {author} {\bibfnamefont {D.~N.}\ \bibnamefont {Baker}}, \bibinfo {author} {\bibfnamefont {H.~E.}\ \bibnamefont {Spence}}, \bibinfo {author} {\bibfnamefont {G.~D.}\ \bibnamefont {Reeves}}, \bibinfo {author} {\bibfnamefont {M.~G.}\ \bibnamefont {Henderson}}, \bibinfo {author} {\bibfnamefont {C.~A.}\ \bibnamefont {Kletzing}}, \bibinfo {author} {\bibfnamefont {W.~S.}\ \bibnamefont {Kurth}}, \bibinfo {author} {\bibfnamefont {G.~B.}\ \bibnamefont {Hospodarsky}}, \bibinfo {author} {\bibfnamefont {J.~B.}\ \bibnamefont {Blake}}, \bibinfo {author} {\bibfnamefont {J.~F.}\ \bibnamefont {Fennell}}, \bibinfo {author} {\bibfnamefont {S.~G.}\ \bibnamefont {Claudepierre}},\ and\ \bibinfo {author} {\bibfnamefont {S.~G.}\ \bibnamefont {Kanekal}},\ }\bibfield  {title} {\bibinfo {title} {Rapid local acceleration of relativistic radiation-belt electrons by magnetospheric chorus},\ }\href {https://doi.org/10.1038/nature12889} {\bibfield  {journal} {\bibinfo  {journal} {Nature}\ }\textbf {\bibinfo {volume} {504}},\ \bibinfo {pages} {411} (\bibinfo {year} {2013})}\BibitemShut {NoStop}%
\bibitem [{\citenamefont {Reeves}\ \emph {et~al.}(2013)\citenamefont {Reeves}, \citenamefont {Spence}, \citenamefont {Henderson}, \citenamefont {Morley}, \citenamefont {Friedel}, \citenamefont {Funsten}, \citenamefont {Baker}, \citenamefont {Kanekal}, \citenamefont {Blake}, \citenamefont {Fennell}, \citenamefont {Claudepierre}, \citenamefont {Thorne}, \citenamefont {Turner}, \citenamefont {Kletzing}, \citenamefont {Kurth}, \citenamefont {Larsen},\ and\ \citenamefont {Niehof}}]{Reeves2013}%
  \BibitemOpen
  \bibfield  {author} {\bibinfo {author} {\bibfnamefont {G.~D.}\ \bibnamefont {Reeves}}, \bibinfo {author} {\bibfnamefont {H.~E.}\ \bibnamefont {Spence}}, \bibinfo {author} {\bibfnamefont {M.~G.}\ \bibnamefont {Henderson}}, \bibinfo {author} {\bibfnamefont {S.~K.}\ \bibnamefont {Morley}}, \bibinfo {author} {\bibfnamefont {R.~H.~W.}\ \bibnamefont {Friedel}}, \bibinfo {author} {\bibfnamefont {H.~O.}\ \bibnamefont {Funsten}}, \bibinfo {author} {\bibfnamefont {D.~N.}\ \bibnamefont {Baker}}, \bibinfo {author} {\bibfnamefont {S.~G.}\ \bibnamefont {Kanekal}}, \bibinfo {author} {\bibfnamefont {J.~B.}\ \bibnamefont {Blake}}, \bibinfo {author} {\bibfnamefont {J.~F.}\ \bibnamefont {Fennell}}, \bibinfo {author} {\bibfnamefont {S.~G.}\ \bibnamefont {Claudepierre}}, \bibinfo {author} {\bibfnamefont {R.~M.}\ \bibnamefont {Thorne}}, \bibinfo {author} {\bibfnamefont {D.~L.}\ \bibnamefont {Turner}}, \bibinfo {author} {\bibfnamefont {C.~A.}\ \bibnamefont {Kletzing}}, \bibinfo {author} {\bibfnamefont {W.~S.}\ \bibnamefont {Kurth}}, \bibinfo {author} {\bibfnamefont {B.~A.}\ \bibnamefont {Larsen}},\ and\ \bibinfo {author} {\bibfnamefont {J.~T.}\ \bibnamefont {Niehof}},\ }\bibfield  {title} {\bibinfo {title} {Electron acceleration in the heart of the {V}an {A}llen radiation belts},\ }\href {https://doi.org/10.1126/science.1237743} {\bibfield  {journal} {\bibinfo  {journal} {Science}\ }\textbf {\bibinfo {volume} {341}},\ \bibinfo {pages} {991} (\bibinfo {year} {2013})}\BibitemShut {NoStop}%
\bibitem [{\citenamefont {Thorne}\ \emph {et~al.}(2010)\citenamefont {Thorne}, \citenamefont {Ni}, \citenamefont {Tao}, \citenamefont {Horne},\ and\ \citenamefont {Meredith}}]{Thorne2010}%
  \BibitemOpen
  \bibfield  {author} {\bibinfo {author} {\bibfnamefont {R.~M.}\ \bibnamefont {Thorne}}, \bibinfo {author} {\bibfnamefont {B.}~\bibnamefont {Ni}}, \bibinfo {author} {\bibfnamefont {X.}~\bibnamefont {Tao}}, \bibinfo {author} {\bibfnamefont {R.~B.}\ \bibnamefont {Horne}},\ and\ \bibinfo {author} {\bibfnamefont {N.~P.}\ \bibnamefont {Meredith}},\ }\bibfield  {title} {\bibinfo {title} {Scattering by chorus waves as the dominant cause of diffuse auroral precipitation},\ }\href {https://doi.org/10.1038/nature09467} {\bibfield  {journal} {\bibinfo  {journal} {Nature}\ }\textbf {\bibinfo {volume} {467}},\ \bibinfo {pages} {943} (\bibinfo {year} {2010})}\BibitemShut {NoStop}%
\bibitem [{\citenamefont {Helliwell}(1967)}]{Helliwell1967}%
  \BibitemOpen
  \bibfield  {author} {\bibinfo {author} {\bibfnamefont {R.~A.}\ \bibnamefont {Helliwell}},\ }\bibfield  {title} {\bibinfo {title} {A theory of discrete {VLF} emissions from the magnetosphere},\ }\href@noop {} {\bibfield  {journal} {\bibinfo  {journal} {J.~Geophys. Res.}\ }\textbf {\bibinfo {volume} {72}},\ \bibinfo {pages} {4773} (\bibinfo {year} {1967})}\BibitemShut {NoStop}%
\bibitem [{\citenamefont {Nunn}(1974)}]{Nunn1974}%
  \BibitemOpen
  \bibfield  {author} {\bibinfo {author} {\bibfnamefont {D.}~\bibnamefont {Nunn}},\ }\bibfield  {title} {\bibinfo {title} {A self-consistent theory of triggered {VLF} emissions},\ }\href {https://doi.org/http://dx.doi.org/10.1016/0032-0633(74)90070-1} {\bibfield  {journal} {\bibinfo  {journal} {Planet. Space Sci.}\ }\textbf {\bibinfo {volume} {22}},\ \bibinfo {pages} {349} (\bibinfo {year} {1974})}\BibitemShut {NoStop}%
\bibitem [{\citenamefont {Trakhtengerts}(1995)}]{Trakhtengerts1995}%
  \BibitemOpen
  \bibfield  {author} {\bibinfo {author} {\bibfnamefont {V.~Y.}\ \bibnamefont {Trakhtengerts}},\ }\bibfield  {title} {\bibinfo {title} {Magnetosphere cyclotron maser: {B}ackward wave oscillator generation regime},\ }\href@noop {} {\bibfield  {journal} {\bibinfo  {journal} {J.~Geophys. Res.}\ }\textbf {\bibinfo {volume} {100}},\ \bibinfo {pages} {17205} (\bibinfo {year} {1995})}\BibitemShut {NoStop}%
\bibitem [{\citenamefont {Omura}\ \emph {et~al.}(2008)\citenamefont {Omura}, \citenamefont {Katoh},\ and\ \citenamefont {Summers}}]{Omura2008}%
  \BibitemOpen
  \bibfield  {author} {\bibinfo {author} {\bibfnamefont {Y.}~\bibnamefont {Omura}}, \bibinfo {author} {\bibfnamefont {Y.}~\bibnamefont {Katoh}},\ and\ \bibinfo {author} {\bibfnamefont {D.}~\bibnamefont {Summers}},\ }\bibfield  {title} {\bibinfo {title} {Theory and simulation of the generation of whistler-mode chorus},\ }\href {https://doi.org/10.1029/2007JA012622} {\bibfield  {journal} {\bibinfo  {journal} {J.~Geophys. Res.}\ }\textbf {\bibinfo {volume} {113}},\ \bibinfo {pages} {A04223} (\bibinfo {year} {2008})}\BibitemShut {NoStop}%
\bibitem [{\citenamefont {Tao}\ \emph {et~al.}(2021)\citenamefont {Tao}, \citenamefont {Zonca},\ and\ \citenamefont {Chen}}]{Tao2021}%
  \BibitemOpen
  \bibfield  {author} {\bibinfo {author} {\bibfnamefont {X.}~\bibnamefont {Tao}}, \bibinfo {author} {\bibfnamefont {F.}~\bibnamefont {Zonca}},\ and\ \bibinfo {author} {\bibfnamefont {L.}~\bibnamefont {Chen}},\ }\bibfield  {title} {\bibinfo {title} {A “{T}rap-{R}elease-{A}mplify” model of chorus waves},\ }\href {https://doi.org/https://doi.org/10.1029/2021JA029585} {\bibfield  {journal} {\bibinfo  {journal} {J.~Geophys. Res. Space Physics}\ }\textbf {\bibinfo {volume} {126}},\ \bibinfo {pages} {e2021JA029585} (\bibinfo {year} {2021})}\BibitemShut {NoStop}%
\bibitem [{\citenamefont {Zonca}\ \emph {et~al.}(2021)\citenamefont {Zonca}, \citenamefont {Tao},\ and\ \citenamefont {Chen}}]{Zonca2021}%
  \BibitemOpen
  \bibfield  {author} {\bibinfo {author} {\bibfnamefont {F.}~\bibnamefont {Zonca}}, \bibinfo {author} {\bibfnamefont {X.}~\bibnamefont {Tao}},\ and\ \bibinfo {author} {\bibfnamefont {L.}~\bibnamefont {Chen}},\ }\bibfield  {title} {\bibinfo {title} {Nonlinear dynamics and phase space transport by chorus emission},\ }\href@noop {} {\bibfield  {journal} {\bibinfo  {journal} {Reviews of Modern Plasma Physics}\ }\textbf {\bibinfo {volume} {5}},\ \bibinfo {pages} {1} (\bibinfo {year} {2021})}\BibitemShut {NoStop}%
\bibitem [{\citenamefont {Zonca}\ \emph {et~al.}(2022)\citenamefont {Zonca}, \citenamefont {Tao},\ and\ \citenamefont {Chen}}]{Zonca2022}%
  \BibitemOpen
  \bibfield  {author} {\bibinfo {author} {\bibfnamefont {F.}~\bibnamefont {Zonca}}, \bibinfo {author} {\bibfnamefont {X.}~\bibnamefont {Tao}},\ and\ \bibinfo {author} {\bibfnamefont {L.}~\bibnamefont {Chen}},\ }\bibfield  {title} {\bibinfo {title} {A theoretical framework of chorus wave excitation},\ }\href {https://doi.org/https://doi.org/10.1029/2021JA029760} {\bibfield  {journal} {\bibinfo  {journal} {J.~Geophys. Res. Space Physics}\ }\textbf {\bibinfo {volume} {127}},\ \bibinfo {pages} {e2021JA029760} (\bibinfo {year} {2022})}\BibitemShut {NoStop}%
\bibitem [{\citenamefont {Bonham}\ and\ \citenamefont {Bhattacharjee}(2025)}]{Bonham2025}%
  \BibitemOpen
  \bibfield  {author} {\bibinfo {author} {\bibfnamefont {B.}~\bibnamefont {Bonham}}\ and\ \bibinfo {author} {\bibfnamefont {A.}~\bibnamefont {Bhattacharjee}},\ }\bibfield  {title} {\bibinfo {title} {Whistler chorus amplification in the magnetosphere: The nonlinear free-electron laser model and the ginzburg-landau equation},\ }\href {https://doi.org/https://doi.org/10.1029/2025GL117547} {\bibfield  {journal} {\bibinfo  {journal} {Geophys. Res. Lett.}\ }\textbf {\bibinfo {volume} {52}},\ \bibinfo {pages} {e2025GL117547} (\bibinfo {year} {2025})}\BibitemShut {NoStop}%
\bibitem [{\citenamefont {Soto-Chavez}\ \emph {et~al.}(2012)\citenamefont {Soto-Chavez}, \citenamefont {Bhattacharjee},\ and\ \citenamefont {Ng}}]{SotoChavez2012}%
  \BibitemOpen
  \bibfield  {author} {\bibinfo {author} {\bibfnamefont {A.~R.}\ \bibnamefont {Soto-Chavez}}, \bibinfo {author} {\bibfnamefont {A.}~\bibnamefont {Bhattacharjee}},\ and\ \bibinfo {author} {\bibfnamefont {C.~S.}\ \bibnamefont {Ng}},\ }\bibfield  {title} {\bibinfo {title} {Chorus wave amplification: {A} free electron laser in the {E}arth's magnetosphere},\ }\href {https://doi.org/http://dx.doi.org/10.1063/1.3676157} {\bibfield  {journal} {\bibinfo  {journal} {Phys. Plasmas}\ }\textbf {\bibinfo {volume} {19}},\ \bibinfo {pages} {010701} (\bibinfo {year} {2012})}\BibitemShut {NoStop}%
\bibitem [{\citenamefont {Angelopoulos}(2008)}]{Angelopoulos2008}%
  \BibitemOpen
  \bibfield  {author} {\bibinfo {author} {\bibfnamefont {V.}~\bibnamefont {Angelopoulos}},\ }\bibfield  {title} {\bibinfo {title} {The {THEMIS} mission},\ }\href {https://doi.org/10.1007/s11214-008-9336-1} {\bibfield  {journal} {\bibinfo  {journal} {Space Sci. Rev.}\ }\textbf {\bibinfo {volume} {141}},\ \bibinfo {pages} {5} (\bibinfo {year} {2008})}\BibitemShut {NoStop}%
\bibitem [{\citenamefont {Neishtadt}\ \emph {et~al.}(2013)\citenamefont {Neishtadt}, \citenamefont {Vasiliev},\ and\ \citenamefont {Artemyev}}]{Neishtadt2013}%
  \BibitemOpen
  \bibfield  {author} {\bibinfo {author} {\bibfnamefont {A.~I.}\ \bibnamefont {Neishtadt}}, \bibinfo {author} {\bibfnamefont {A.~A.}\ \bibnamefont {Vasiliev}},\ and\ \bibinfo {author} {\bibfnamefont {A.~V.}\ \bibnamefont {Artemyev}},\ }\bibfield  {title} {\bibinfo {title} {Capture into resonance and escape from it in a forced nonlinear pendulum},\ }\href {https://doi.org/10.1134/S1560354713060087} {\bibfield  {journal} {\bibinfo  {journal} {Regular and Chaotic Dynamics}\ }\textbf {\bibinfo {volume} {18}},\ \bibinfo {pages} {686} (\bibinfo {year} {2013})}\BibitemShut {NoStop}%
\bibitem [{\citenamefont {Artemyev}\ \emph {et~al.}(2023)\citenamefont {Artemyev}, \citenamefont {Albert}, \citenamefont {Neishtadt},\ and\ \citenamefont {Mourenas}}]{Artemyev2023}%
  \BibitemOpen
  \bibfield  {author} {\bibinfo {author} {\bibfnamefont {A.~V.}\ \bibnamefont {Artemyev}}, \bibinfo {author} {\bibfnamefont {J.~M.}\ \bibnamefont {Albert}}, \bibinfo {author} {\bibfnamefont {A.~I.}\ \bibnamefont {Neishtadt}},\ and\ \bibinfo {author} {\bibfnamefont {D.}~\bibnamefont {Mourenas}},\ }\bibfield  {title} {\bibinfo {title} {The effect of wave frequency drift on the electron nonlinear resonant interaction with whistler-mode waves},\ }\href {https://doi.org/10.1063/5.0131297} {\bibfield  {journal} {\bibinfo  {journal} {Physics of Plasmas}\ }\textbf {\bibinfo {volume} {30}},\ \bibinfo {pages} {012901} (\bibinfo {year} {2023})}\BibitemShut {NoStop}%
\bibitem [{\citenamefont {Tao}(2014)}]{Tao2014b}%
  \BibitemOpen
  \bibfield  {author} {\bibinfo {author} {\bibfnamefont {X.}~\bibnamefont {Tao}},\ }\bibfield  {title} {\bibinfo {title} {A numerical study of chorus generation and the related variation of wave intensity using the {DAWN} code},\ }\href {https://doi.org/10.1002/2014JA019820} {\bibfield  {journal} {\bibinfo  {journal} {J.~Geophys. Res. Space Physics}\ }\textbf {\bibinfo {volume} {119}},\ \bibinfo {pages} {3362} (\bibinfo {year} {2014})}\BibitemShut {NoStop}%
\bibitem [{\citenamefont {Lichtenberg}\ and\ \citenamefont {Lieberman}(1983)}]{Lichtenberg1983}%
  \BibitemOpen
  \bibfield  {author} {\bibinfo {author} {\bibfnamefont {A.}~\bibnamefont {Lichtenberg}}\ and\ \bibinfo {author} {\bibfnamefont {M.}~\bibnamefont {Lieberman}},\ }\href@noop {} {\emph {\bibinfo {title} {Regular and Chaotic Dynamics}}},\ \bibinfo {edition} {2nd}\ ed.\ (\bibinfo  {publisher} {Springer Verlag},\ \bibinfo {address} {New York},\ \bibinfo {year} {1983})\BibitemShut {NoStop}%
\bibitem [{\citenamefont {Artemyev}\ \emph {et~al.}(2018)\citenamefont {Artemyev}, \citenamefont {Neishtadt}, \citenamefont {Vainchtein}, \citenamefont {Vasiliev}, \citenamefont {Vasko},\ and\ \citenamefont {Zelenyi}}]{Artemyev2018}%
  \BibitemOpen
  \bibfield  {author} {\bibinfo {author} {\bibfnamefont {A.}~\bibnamefont {Artemyev}}, \bibinfo {author} {\bibfnamefont {A.}~\bibnamefont {Neishtadt}}, \bibinfo {author} {\bibfnamefont {D.}~\bibnamefont {Vainchtein}}, \bibinfo {author} {\bibfnamefont {A.}~\bibnamefont {Vasiliev}}, \bibinfo {author} {\bibfnamefont {I.}~\bibnamefont {Vasko}},\ and\ \bibinfo {author} {\bibfnamefont {L.}~\bibnamefont {Zelenyi}},\ }\bibfield  {title} {\bibinfo {title} {Trapping (capture) into resonance and scattering on resonance: {S}ummary of results for space plasma systems},\ }\href {https://doi.org/https://doi.org/10.1016/j.cnsns.2018.05.004} {\bibfield  {journal} {\bibinfo  {journal} {Communications in Nonlinear Science and Numerical Simulation}\ }\textbf {\bibinfo {volume} {65}},\ \bibinfo {pages} {111} (\bibinfo {year} {2018})}\BibitemShut {NoStop}%
\bibitem [{\citenamefont {Wu}\ \emph {et~al.}(2023)\citenamefont {Wu}, \citenamefont {Huang}, \citenamefont {Wu}, \citenamefont {Tao}, \citenamefont {Katoh},\ and\ \citenamefont {Wang}}]{Wu2023}%
  \BibitemOpen
  \bibfield  {author} {\bibinfo {author} {\bibfnamefont {Z.}~\bibnamefont {Wu}}, \bibinfo {author} {\bibfnamefont {H.}~\bibnamefont {Huang}}, \bibinfo {author} {\bibfnamefont {Y.}~\bibnamefont {Wu}}, \bibinfo {author} {\bibfnamefont {X.}~\bibnamefont {Tao}}, \bibinfo {author} {\bibfnamefont {Y.}~\bibnamefont {Katoh}},\ and\ \bibinfo {author} {\bibfnamefont {X.}~\bibnamefont {Wang}},\ }\bibfield  {title} {\bibinfo {title} {Connection between chorus wave amplitude and background magnetic field inhomogeneity: {A} parametric study},\ }\href {https://doi.org/https://doi.org/10.1029/2023GL106397} {\bibfield  {journal} {\bibinfo  {journal} {Geophys. Res. Lett.}\ }\textbf {\bibinfo {volume} {50}},\ \bibinfo {pages} {e2023GL106397} (\bibinfo {year} {2023})}\BibitemShut {NoStop}%
\bibitem [{\citenamefont {Tao}\ \emph {et~al.}(2017)\citenamefont {Tao}, \citenamefont {Zonca},\ and\ \citenamefont {Chen}}]{Tao2017b}%
  \BibitemOpen
  \bibfield  {author} {\bibinfo {author} {\bibfnamefont {X.}~\bibnamefont {Tao}}, \bibinfo {author} {\bibfnamefont {F.}~\bibnamefont {Zonca}},\ and\ \bibinfo {author} {\bibfnamefont {L.}~\bibnamefont {Chen}},\ }\bibfield  {title} {\bibinfo {title} {Investigations of the electron phase space dynamics in triggered whistler wave emissions using low noise $\delta f$ method},\ }\href {https://doi.org/10.1088/1361-6587/aa759a} {\bibfield  {journal} {\bibinfo  {journal} {Plasma Phys. Controlled Fusion}\ }\textbf {\bibinfo {volume} {59}},\ \bibinfo {pages} {094001} (\bibinfo {year} {2017})}\BibitemShut {NoStop}%
\end{thebibliography}

%apsrev4-2.bst 2019-01-14 (MD) hand-edited version of apsrev4-1.bst
%Control: key (0)
%Control: author (8) initials jnrlst
%Control: editor formatted (1) identically to author
%Control: production of article title (0) allowed
%Control: page (0) single
%Control: year (1) truncated
%Control: production of eprint (0) enabled
%

% \section*{Data availability}
% The data supporting this study can be obtained by contacting the corresponding
% author.

\begin{acknowledgments}
  This work was supported by NSFC Grant No. 42474218. 
\end{acknowledgments}

\section*{Conflict of Interest}
The authors declare no conflicts of interest.

\end{document}